# Disorder-immune metasurfaces with constituents exhibiting the anapole mode

Hao Song[1], Neng Wang[1,*], Kuai Yu[1], Jihong Pei[2] and Guo Ping Wang[1,*]

[1] *Institute of Microscale Optoelectronics, Shenzhen University, Shenzhen 518060, P. R. China*

[2] *College of Electronics and Information Engineering, Shenzhen University, Shenzhen 518060,* P. R. *China*

*Corresponding author:* nwang17@szu.edu.cn, gpwang@szu.edu.cn

**Abstract**

Common optical metasurfaces are 2-dimensional functional devices composed of periodically arranged subwavelength constituents. Here, we achieved the positional-disorder-immune metasurfaces composed of core-shell cylinders which successively exhibit the magnetic dipole (MD) resonant, non-radiating anapole, and electric dipole (ED) resonant modes when their outer radii are fixed and the inner radii changes continuously in a range. The performances of the metasurfaces under a periodically structural design are not degraded even when the positions of the cylinders are subjected to random and considerable displacements. The positional-disorder-immunity is due to the weak non-local effect of the metasurfaces. Because the multiple scattering among cylinders is weak and insensitive to the spacing among the cylinders around the ED and MD resonant modes and vanishing irrespective of the spacing at the non-radiating anapole mode, the reflection properties including the reflection phase and reflectivity of the metasurfaces are insensitive to the spacing between neighboring cylinders for this entire variation range of the inner radius. The positional-disorder-immunity make the metasurface to be adapted to some unstable and harsh environments.

**I. Introduction**

Metasurfaces can tailor the wavefront of an electromagnetic (EM) wave into an arbitrarily desired shape by adjusting the geometrical and/or constitutive parameters of the subwavelength optical scatters [1], thereby achieving fruitful applications such as the optical vortex generation [2], subwavelength focusing [3], carpet cloak [4], wavefront deflection [1], holographic imaging [5], perfect absorption [6], beam splitting [7] and so on. The high efficacy on the wavefront manipulation is benefited from the abrupt phase changes between the two sides of the subwavelength scatter. Apart from the Pancharatnam-Berry phase method [8], the Mie resonant excitations of low-loss dielectric scatters provide another route to realize the phase modulation [9]. Moreover, the rapid changes around the resonant peak enable the flexible control of both the magnitude and phase of the scattered EM wave [10-12]. However, to implement a full $2\pi$ phase modulation based on a single resonant mode excitation is impossible because the range of phase change is limited to $\pi$ for the Lorentzian resonators [13]. Therefore, a covering of multiple resonant mode excitations is necessary for the full $2\pi$ phase modulation.

Most metasurfaces are in periodic structures [14-21]. However, the periodicity and spatial symmetry of the structure lead a more pronounced spatial dispersion [22], and largely reduces the degree of freedom of the structure to control the scattering and transporting of the EM wave in a more flexible way, compared with the disordered metasurfaces [23-31]. And the perfectly periodic structures require much more sophisticated and precise fabrication technologies such as nano-lithographic techniques, which are time-consuming and comparatively expensive [32, 33]. Moreover, the performances of the metasurface with a periodic structure are usually sensitive to the spacing among the neighboring subwavelength constituents [34-37]. This sensitivity leads the metasurface cannot deal with the positional disorders induced by some inevitable external forces in unstable and harsh environments.

Resonant modes can localize the scattered field near the surface of the scatter, thereby making the scattered field intensities decreasing dramatically far away from the surface. The weak multiple scattering among the scatters due to the weak scattered field away from the surfaces render the EM wave properties of the whole structure to less depend on the spacing among the scatters when they are not so close to each other [38]. As the extremest case, the anapole mode [39-44] which is pure non-radiative can even localize the scattered field totally inside the scatter. The non-radiating behavior of the anapole mode provides a new mechanism for achieving the EM wave transparency [41, 45]. Because the light is grasped inside the scatter without leaking, light-matter interaction inside the scatter can be enhanced to enable various applications of the anapole mode, such as the nonlinear effect enhancement [46, 47], nanolasers [48] and sensing [49]. Since the scattered field outside the scatters are vanishing, the multiple scattering among the scatters for the anapole mode excitation is absent irrespective of their spacing. As such, the EM wave properties of the whole structure will not depend on how the scatters which are exhibiting the anapole mode are arranged.

In this work, we achieved the metasurfaces with performances immune to a large positional disorder using one-dimensional arrays of the metallic core- high dielectric shell cylinders. For a fixed outer radius, the magnetic dipole (MD) resonant, anapole, and electric dipole (ED) resonant modes of the cylinder are excited successively as the ratio of inner radius to outer radius $α$ increases from 0.20 to 0.62. Because both the ED and MD resonant modes are involved, the reflection phase can be modulated from $-π$ to $π$ in this inner radius range. Around the ED or MD resonance, the modulated phase is insensitive to the spacing among the adjacent cylinders. More importantly, the modulated phase is completely independent of the spacing at the anapole mode. As a consequence, the phase modulation for this entire range of $α$ is almost independent of the spacing. As such, the performances of the metasurfaces based on the phase modulation are robust against a large positional disorder of the cylinders. In some unstable and harsh circumstances, such as the outdoors under the sunshine and the battleground, positional disorders are inevitably induced, such as the local thermal expansion or cold contraction and displacements due to the mechanical impacts. Therefore, the positional-disorder-immunity will promote and improve the applications of the metasurfaces in such kind of circumstances.

## II. Scattering and multiple scattering properties of the core-shell cylinders

The metallic core- high dielectric shell cylinder we considered in this paper is shown in Fig. 1(a),

which has an inner radius $r_0$ and an outer radius $r_1$. The shell is made of LiTaO3 with refractive index $m$ and the surrounding is air. Throughout this paper, we consider that the outer radius of the cylinder is $k_0 r_1 = 0.98$, where $k_0$ is the wavenumber in the surrounding. For convenience, we define a dimensionless parameter $\alpha = r_0 / r_1$ to characterize the inner radius. The polarization of the external plane wave is parallel to the central line of the cylinder (along the $z$ direction) and the wave vector is perpendicular to the central line (along the negative-$y$ direction). At the incident frequency 0.38 THz, the refractive index of LiTaO3 is $m = 6.44 + 0.0069i$ [50]. According to the Mie theory for core-shell scatters [51], the scattering, extinction, and absorption efficiencies are obtained as:

$$N_{sca} = \frac{2}{x}\left[|b_0|^2 + 2\sum_{n=1}^{\infty}|b_n|^2\right], \quad (1)$$

$$N_{ext} = \frac{2}{x}\text{Re}\left\{b_0 + 2\sum_{n=1}^{\infty}b_n\right\}, \quad (2)$$

$$N_{abs} = N_{ext} - N_{sca}, \quad (3)$$

where $x = k_0 r_1$ is the dimensionless size parameter in the surrounding, and $b_n$ are the Mie coefficients of the cylinder expressed as [51, 52]

$$b_n = \frac{A_n + mB_n}{C_n + mD_n}. \quad (4)$$

Here the auxiliary functions are defined as

$$A_n = J_n(k_1 r_0)Y_n(k_1 r_1)J_n'(k_0 r_1) - J_n(k_1 r_1)Y_n(k_1 r_0)J_n'(k_0 r_1),$$

$$B_n = J_n(k_0 r_1)Y_n(k_1 r_0)J_n'(k_1 r_1) - J_n(k_1 r_0)J_n(k_0 r_1)Y_n'(k_1 r_1),$$

$$C_n = J_n(k_1 r_1)Y_n(k_1 r_0)H_n'(k_0 r_1) - J_n(k_1 r_0)Y_n(k_1 r_1)H_n'(k_0 r_1), \quad (5)$$

$$D_n = J_n(k_1 r_0)H_n(k_0 r_1)Y_n'(k_1 r_1) - Y_n(k_1 r_0)H_n(k_0 r_1)J_n'(k_1 r_1),$$

where $J_n$, $Y_n$, and $H_n$ are, respectively, the Bessel function of the first kind, the Bessel function of the second kind, and the Hankel function of the first kind, and $k_1 = mk_0$ is the wavenumber inside the shell.

In Fig. 1(b) we showed the scattering efficiency (SE) and absorption efficiency (AE) based on Eqs. (1)-(5) as functions of $\alpha$ by the red and blue solid lines, respectively. The vanishing small values of the AE curve indicates that the absorption of a single core-shell cylinder can be safely neglected for most cases. In Fig. 1(c), we showed $2|b_0|/x$, $4|b_1|/x$, $4|b_3|/x$ which are corresponding to the contributions from the electric dipole (ED), magnetic dipole (MD), electric quadrupole (EQ) modes, respectively [53]. The contributions of Mie coefficients of other orders for SE are so weak that are ignored. There are two peaks (points A and D) and one dip (point C) for the ED mode, one peak (point B) and one dip (point C) for the MD mode, and one peak (point E) for the EQ mode when $\alpha$ ranges from 0.01 to 0.62. The peaks of the modes correspond to the resonances of the modes. It is interesting that the ED and MD dips meet at the same point (point C) and approaches to zero, meanwhile, the EQ mode is vanishing small at this point. Therefore, the non-radiating anapole mode is excited at the point C [42, 54].

In Figs. 1(d)-1(f), we showed the normalized scattering electric field intensity distributions $|E_{sz}|$ at the points B (MD resonant mode), C (anapole mode), and D (ED resonant mode), respectively. For the MD and ED resonant modes, the scattering fields are localized near the surface of the cylinders, see Figs. 1(d) and 1(f), while for the anapole mode, there is no scattering field outside the cylinder, see Fig. 1(e). The scattering fields outside the cylinders will determine the strength of the multiple scattering among multiple cylinders. To the zeroth order approximation, the multiple scattering strength is proportional to $\mathbf{E}_{is}(\mathbf{r}_j - \mathbf{r}_i)$, where $\mathbf{E}_{is}$ is the scattering field of the $i$th cylinder, $\mathbf{r}_i, \mathbf{r}_j$ are positions of the $i$th and $j$th cylinders. Therefore, for the anapole mode excitation, the multiple scattering among the cylinders is vanishing. In Figs. 1(g)-(i), we plotted the normalized scattering field distributions for three identical cylinders for the MD resonant, anapole, and ED resonant modes, respectively. The cylinders are aligned along $x$ direction with the same spacing. We can see that for the anapole excitation, the field distribution keeps invariant, while for the MD and ED resonant mode excitations, the field distributions are changed when other cylinders are close.

**III. Structural parameters-dependent reflection properties of the metasurfaces**

We designed the reflective metasurfaces by arranging the core-shell cylinders with the fixed outer radii $r_1$ and well chosen inner core radii $r_0$ along the $x$ direction on the metallic film, as schematically shown in Fig. 1(a).

We first investigated the reflection properties of the metasurface composed of identical cylinders arranged periodically. Figs. 2(a) and 2(b), respectively, show the reflection phases and reflectivities as functions of $\alpha$ when it changes from 0.20 to 0.62 continuously for different spacing $d$. This range of $\alpha$ covers the points B, C, D, and E in Fig. 1(c). The spacing $d$ is defined as the nearest distance between the surfaces of neighboring cylinders, see Fig. 1(a). The reflection phase is defined as the phase difference between the reflection and incident waves at the top of the cylinders. It can be seen that the reflection phases cover the full $2\pi$ range (from $-\pi$ to $\pi$) and the reflectivities are close to unity for all spacing ranging from $0.04\lambda$ to $0.14\lambda$. It is worth noticing that for $\alpha = 0.26$ which just corresponds to the anapole excitation, the reflection phases are invariant for all spacing. This is expected because the cylinder array is transparent for the EM wave no matter what the concentration is due to the non-radiating properties of the anapole mode. Also, the reflectivity find dips at $\alpha = 0.26$. This is also expected since the absorption is enhanced as the strong EM fields are localized inside the absorptive shell when the anapole mode is excited. The depths of the reflectivity dips are just proportional to the concentrations of the arrays. What's more, for the MD and ED resonant mode excitations, because the scattering fields are decaying fast especially along the $x$ direction, see Figs. 1(d) and 1(f), the multiple scatterings among the cylinders are very weak and insensitive to the spacing when they are not so close to each other. Consequently, for this entire range of $\alpha$, both the reflection phases and reflectivities are almost the same for all spacing ranging from 0.04λ to 0.14λ, see Figs. 2(a) and 2(b).

As a comparison, we also considered the case when $\alpha$ changes from 0.01 to 0.20. This range of $\alpha$ also involves the transition from the ED resonant mode to the MD resonant mode. The reflection phases and reflectivities as functions of $\alpha$ for different spacing $d$ are shown in Figs. 2(c) and 2(d), respectively. We can see the reflection phase also covers the full $2\pi$ range. However, the reflectivities deviate from 1 for $0.10<\alpha<0.15$. This is because the backward scattering is reduced and the EM fields are more likely coupled inside the cylinders when the ED and MD moments are almost equal [55-57], see Fig. 1(c). What's more important, because no anapole mode is excited during this range of $\alpha$, the reflection phases are shifted evidently when the spacing $d$ changes except at $\alpha = 0.055$ and $\alpha = 0.20$ where the ED and MD resonant modes are excited.

**IV. Positional-disorder-immune metasurfaces**

**A. Magnetic mirrors**

In this section, we first considered the use of the metasurfaces as magnetic mirrors [13-18, 34, 58]. One key identity of the magnetic mirror is the reflection phase $|\Delta\phi_E|<\pi/2$, especially, for the perfect magnetic mirrors the reflection phase is zero $\Delta\phi_E = 0$ [20, 21, 59]. Due to the constructive interference, the electric field enhancement at the interfaces makes the magnetic mirrors to be significant for various applications such as molecular fluorescence [19], perfect absorbers [60], subwavelength imaging [15], photocurrent generation and surface enhanced Raman spectroscopy (SERS) [61], and retroreflectors [18].

The spacing between adjacent cylinders in the magnetic mirror is first assumed to be equal. For the spacing $d = 0.09\lambda$, the perfect magnetic mirror with $\Delta\phi_E = 0$ is achieved when $\alpha = 0.27$ according to Fig. 2(a). The corresponding intensity distribution of the total electric field is shown in Fig. 3(a), where the constructive interference at the level just above the cylinders is clearly seen. Keeping $\alpha = 0.27$ invariant, we changed the spacing $d$ from 0.04λ to 0.14λ. The reflection phases and the reflectivities for different spacing are shown by red and blue solid lines in Fig. 3(b), respectively. We can see that the reflection phases are approaching zero and the reflectivities are around 0.85 for any spacing $d$ ranging from 0.04λ to 0.14λ.

We further designed the magnetic mirror composed of the cylinders with a large positional disorder. The spacing between neighboring cylinders are randomly distributed inside the range 0.04λ~0.14λ. Therefore, the positional disorder reaches about 55.56% for the spacing. 10 samples of disordered magnetic mirrors are considered with their reflectivities and averaged reflection phases (the reflection phase is spatially averaged along $x$ direction) are shown by the blue triangles and red circles in Fig. 3(c). It is clearly seen that the reflectivities as well as the averaged reflection phases of all samples are almost the same as those of the ordered magnetic mirror (magnetic mirror with equal spacing). The normalized electric field intensity distribution of a disordered sample is shown in Fig. 3(d). We can see the constructive interference at the level just above the cylinders. The slight inhomogeneity of electric field intensity along $x$ direction is mainly due to the nonuniform concentrations of the cylinders along $x$ direction. Therefore, the perfect magnetic mirrors composed of the cylinders with $\alpha = 0.27$ are robust against a very large positional disorder.

The magnetic mirror with zero reflection phase can be also realized using cylinders with $\alpha = 0.12$ and equal spacing $d = 0.09\lambda$, according to Fig. 2(c). In Fig. 4(a), we plotted the corresponding normalized electric field intensity distribution. The electric field at the top of the cylinder is enhanced due to constructive interference. The inhomogeneity of electric field intensity along $x$ direction at the level just above the cylinders is due to the strong scattering of the cylinders. However, when the spacing deviates from $0.09\lambda$, the reflection phase will be no longer zero and the reflectivity is changed dramatically, as shown in Fig. 4(b). Therefore, if the positional disorder is introduced, the reflection phase will become nonzero and irregular, as shown in Fig. 4(c). Because of the nonuniform reflection phases and reflectivities along the $x$ direction, the reflected wave behaves no longer like a plane wave, see Fig. 4(d).

**B. Reflected wavefront deflection**

The second application of the metasurface we discussed here is the reflected wavefront deflection [1, 62, 63]. Metasurfaces can reshape wavefronts through the phase discontinuities. Moreover, according to the generalized Snell's law [1], a linear phase variation along the metasurface can induce anomalous reflection and transmission. As the cornerstone of the wavefront deflection theory, the generalized reflection law is expressed as

$$n_0 k_0 (\sin\theta_r - \sin\theta_0) = \frac{d\phi}{dx}, \tag{6}$$

where $d\phi/dx$ is the phase gradient along the $x$ direction, $\theta_0$ and $\theta_r$ are the incident and reflection angles. Without loss of generality, we considered the incident angle to be $\theta_0 = 0$ and the reflection angle to be $\theta_r = 20°$, respectively. The metasurface we designed contains 51 cylinders with $\alpha$ ranging from 0.20 to 0.62. We first consider that all the spacing $d$ between adjacent cylinders are identical and are set as $0.09\lambda$. The inner core radii of the cylinders at different positions are designed according to Eq. (6). In Fig. 5(a), we showed the angle dependent reflection intensity $S(\theta)$ in the far field region which is defined as

$$S(\theta) = \lim_{k_0 r \to \infty} |\mathbf{E}_r(\theta)|/E_0, \tag{7}$$

where $\mathbf{E}_r$ is the reflection field, $E_0$ is the amplitude of the incident wave, $(r, \theta)$ is the polar coordinate with an origin at the center of the metasurface. We can see $S(\theta)$ reaches the maximum at $110°$ and drop fast when away from $110°$, indicating a $20°$ wavefront deflection. In Fig. 5(b), we also plotted the reflection field distribution which shows a directional transporting along the direction of $110°$ clearly.

We then introduced the positional disorder to the metasurface. Similarly, the spacing between adjacent cylinders are no longer equal but randomly chosen in the range of $0.04\lambda \sim 0.14\lambda$, keeping the core radii of the cylinders invariant. The reflection intensity $S(\theta)$ for 10 disordered samples are shown in Fig. 5(c), from which we can see the reflection features for all the disordered metasurfaces are almost the same as that of the ordered metasurface, see Fig. 5(a). And as an example, the reflection field distribution of a disordered metasurface shown in Fig. 5(d) is almost the same as that of the ordered metasurface shown in Fig. 5(b). To quantitatively show the subtle differences on the wavefront deflection effects between the ordered and disordered metasurfaces, we depicted the relative difference on the peak values of the reflection intensities, $\delta S_n$, between the ordered and disordered metasurfaces and the angle expansions $\Delta \theta_n$ of the main lobes of $S(\theta)$ of the disordered metasurfaces, in Fig. 5(e). The angle expansion of $S(\theta)$ of the ordered metasurface is shown by the blue solid line. The main lobes of $S(\theta)$ are from $\sim 100°$ to $\sim 120°$, as we can see from Figs. 5(a) and 5(c). The angle expansion can be calculated as $\Delta\theta = \sqrt{\int (\theta-\theta_0)^2 S(\theta)d\theta / \int S(\theta)d\theta}$, where $\theta_0 = 110°$. We can see the fluctuation on peak values does not exceed $\pm 2\%$. Also, the angle expansions for the disordered metasurfaces are fluctuating around $\sim 3.50°$ for the ordered metasurface not exceeding $0.10°$, indicating the unidirectionality of the reflection waves for the ordered and disordered metasurfaces are almost the same.

As a comparison, we also investigated the metasurfaces composed of cylinders with $\alpha$ ranging from 0.01 to 0.20. The core radii of the cylinders on different positions are designed according to Eq. (7) under the condition that the cylinders are equally spaced by $d = 0.09\lambda$. The reflection intensity $S(\theta)$ and the reflection field distribution of the ordered metasurface are shown in Figs. 6(a) and 6(b), respectively, showing a directional transporting of the reflection wave along the direction of $110°$. However, when the positional disorder is introduced, because the reflection phase and reflectivity are so sensitive to the spacing, as can be seen from Figs. 2(c) and 2(d), that the reflection features are changed, see Fig. 6(c), and the wavefronts become curved, see Fig. 6(d). The relative difference on the peak values of the reflection intensities, $\delta S_n$, and the angle expansions $\Delta\theta_n$ are shown in Fig. 6(e) by the red circle line and blue star line, respectively. We can see the reflection intensity along the direction of $110°$ is decreased by about $10\%$. This is expected since reflection along other directions arises when the positional disorder is introduced, see Fig. 6(c). Moreover, the angle expansions are always increasing (from $\sim 3.40°$ to $\sim 3.80°$). Therefore, the unidirectionality is reduced.

**C. Carpet cloak**

The last application of the metasurface considered in this work is the carpet cloak [4, 64, 65]. Covering a bump with a specially designed metasurface can suppress the unwanted scattering from the bump and make it hided from an external observer. Here, our goal is to create an ultrathin and simple carpet cloak steering the reflection beam into the normally incident beam direction. The metallic polygonal bump to be cloaked is symmetric about its central vertical line and has two tilt angles $\theta_1 = 10°$ and $\theta_2 = 20°$ as schematically shown in Fig. 7(a). For the normal incidence, to compensate the phase shift caused by the height difference $h(x)$, the reflection phase at $x$ should be [4, 66]

$$\Delta\phi_E = -2k_0 h(x). \tag{8}$$

The bottom of the bump has a width of $20\lambda$. To implement the metasurface, 51 cylinders are put along the surface of the bump with the equal horizontal spacing $d = 0.09\lambda$, as shown in Fig. 7(b). The core radii of the cylinders are chosen based on Eq. (8) in conjunction with the data of Fig. 2(a). The normalized reflection intensity $S(\theta)$ under the normal incidence for the bump without and with the metasurface covered are shown in Fig. 7(c) by the black and red solid lines, respectively. When the metasurface is absent, diffuse reflection occurs. But when the metasurface is used, the normal reflection is enhanced while reflections along other directions are almost totally suppressed. Therefore, the bump with the metasurface covered behaves like air for the normally incident wave. This is also verified by the total electric field distributions shown in Figs. 7(d) and 7(e).

Consider the positional disorder is introduced that the horizontal spacing between adjacent cylinders are no longer equal but randomly chosen in the range from $0.04\lambda$ to $0.14\lambda$. The core radii of the cylinders are not changed. The normalized reflection intensity $S(\theta)$ for 10 samples are shown in Fig. 7 (f). We can see for all disordered samples, the reflection features are almost the same as that of the ordered case. The incident wave is almost totally reflected along the normal direction, see Fig. 7(g). As a consequence, the carpet cloak is valid even the large positional disorder is introduced.

We also explored the metasurface composed of cylinders with $\alpha$ ranging from 0.01 to 0.20. The core radii of the cylinders are designed according to Eq. (9) considering that the cylinders are equally spaced with $d = 0.09\lambda$. However, because the reflectivities deviate from unity and are sensitive to $\alpha$, see Fig. 2(d), although the normal reflection is still the strongest, remarkable reflection intensities along other directions cannot be avoided, as shown in Fig. 8(a). Therefore, the reflection wave is no longer plane wave-like and the bump cannot be cloaked by the metasurface, see Fig. 8(b). These are also true for the disordered metasurfaces, as shown by Figs. 8(c) and 8(d).

## V. Conclusions

In summary, we have achieved the metasurfaces with performances immune to large positional disorders by using the core-shell cylinders. The core-shell cylinders we used here will exhibit the MD resonant, non-radiating anapole, and ED resonant modes successively when their outer radii are fixed while the inner radii increase continuously, modulating the reflection phase of the metasurface from $-\pi$ to $\pi$. The positional-disorder-immunity of the metasurface is due to the insensitivity of the phase modulation to the changing of spacing between neighboring cylinders. We have explored three kinds of metasurfaces for three different applications, the magnetic mirror, reflected wavefront deflection, and the carpet cloak. The inner radii of the cylinders of the metasurfaces are designed individually assuming that the cylinders are equally spaced. The numerical results reveal that the performances of the metasurfaces are not degraded even when a large positional disorder is introduced, verifying the positional-disorder-immunity of the metasurfaces. For comparison, we have also investigated the metasurfaces composed of core-shell cylinders which exhibit the ED and MD resonant modes only during the full $2\pi$ range phase modulation. Although the phase modulations at the ED and MD resonant modes are almost invariant about the spacing between the neighboring cylinders, they are shifted evidently as the spacing changes when away from the resonant modes. Therefore, the performances of the metasurfaces are degraded remarkably when the positional disorders are introduced. This indicates that the existence of the anapole mode excitation in the whole phase modulation range is key important for the positional-disorder-immunity of the metasurfaces we proposed here. We note that the core-shell cylinders can be replaced by rectangular shaped homogeneous cylinders which can also support the anapole modes by tuning the geometrical parameters to release the fabrication technique in the experiments. The positional-disorder-immune metasurfaces may find applications in unstable and harsh environments where positional disorders are usually inevitable. We also hope our work could promote the study on the metasurfaces with immunity against various disorders.

## Acknowledgements

This work was supported by the National Natural Science Foundation of China (NSFC) through Nos. 11574218, 11734012 and 11904237 and Science and Technology Project of Guangdong through No. 2020B010190001.

**Figures**

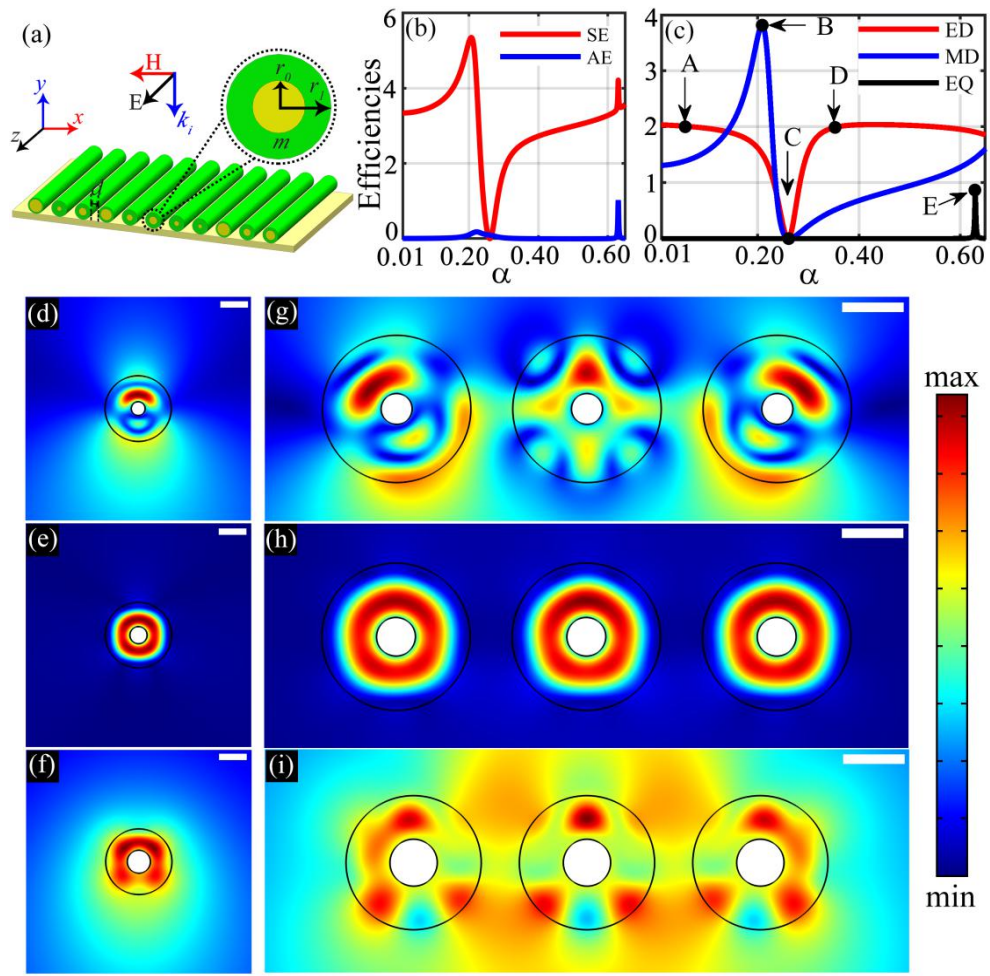

Fig.1 (a) The coordinate settlement and the geometrical sketches of the planar metasurface and core-shell cylinder. The inner and outer radii of the cylinder are $r_0$ and $r_1$. The spacing between the neighboring cylinders is defined as the nearest distance between the neighboring cylinders' surfaces and noted as $d$. (b) SE (red) and AE (blue) of the core-shell cylinder as functions of the ratio $\alpha$. (c) The contributions from different dominate modes. (d)-(f) The normalized scattering field intensity distributions. (d) and (g) the intensity distributions for a single and three identical cylinders at the MD resonant mode excitation corresponding to the point B, (e) and (h) at the anapole mode excitation corresponding to the point C, (f) and (i) at the ED resonant mode excitation corresponding to the point D. The length of thewhite bars is 100 μm.

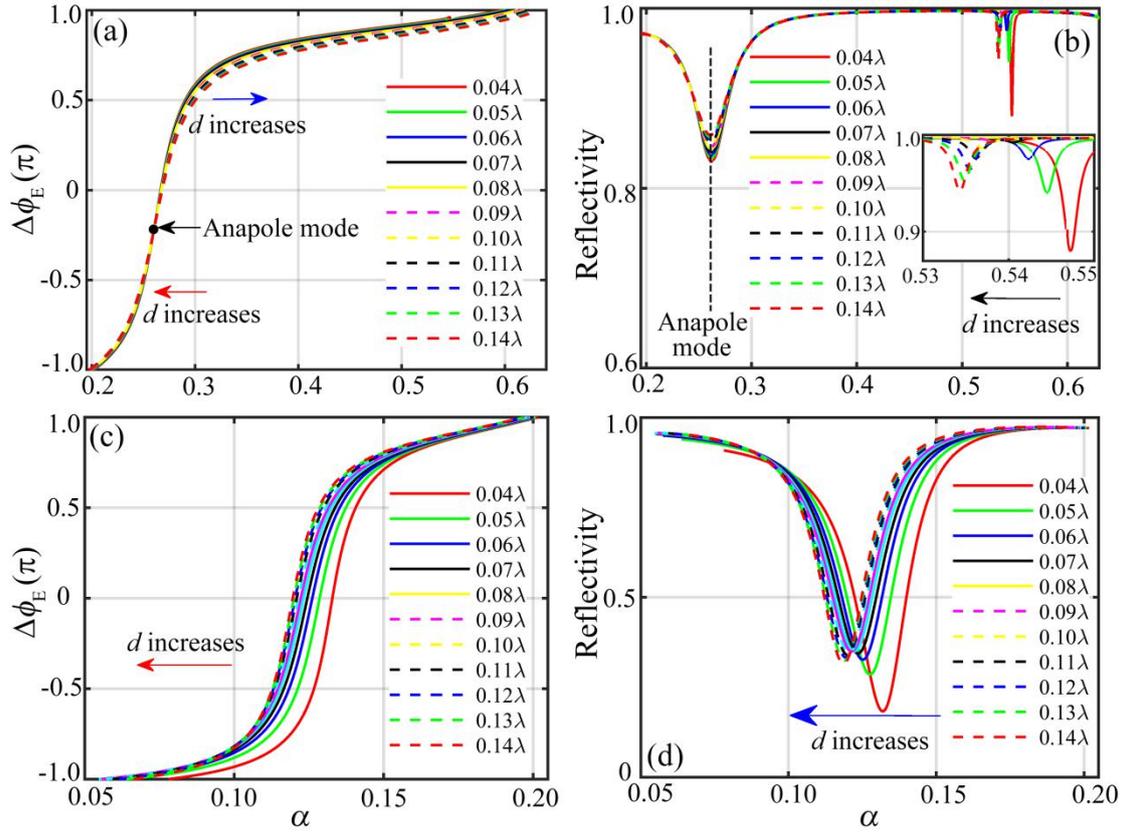

Fig. 2 (a) [(c)] The reflection phases $\Delta\phi_E$ and (b) [(d)] the reflectivities as functions of the ratio $\alpha$ ranging from 0.20 to 0.62 (0.01 to 0.20) for different spacing $d$. The results are calculated assuming the cylinders are equally spaced. The arrows in the figures denote the shifting directions of the curves when the period $d$ increases.

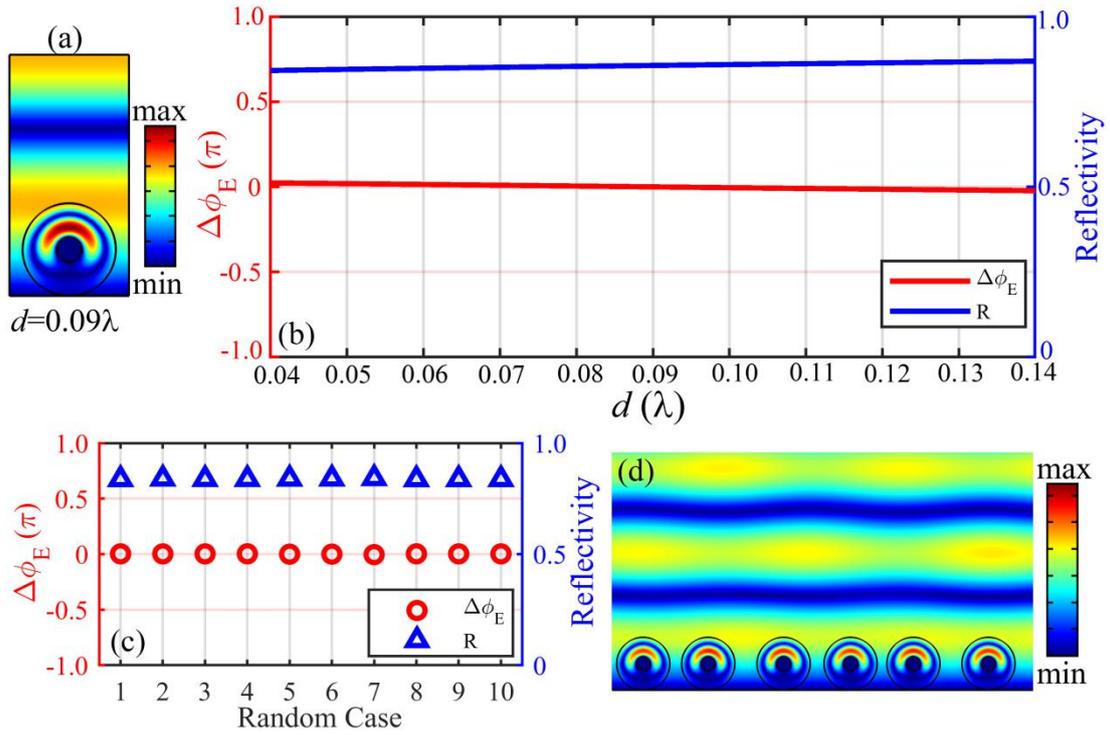

Fig. 3 (a) The distribution of normalized total electric field in a period. The magnetic mirror is in a periodic structure with the spacing between the neighboring cylinders $d = 0.09\lambda$. The ratio of inner radius to outer radius is $\alpha=0.27$. (b) The reflection phase $\Delta\phi_E$ (red solid line) and reflectivity (blue solid line) of the magnetic mirrors in periodic structures as functions of the spacing $d$. (c) The reflectivities (blue triangles) and averaged reflection phases for different disordered samples. The averaged reflection phase is obtained by spatially averaging the reflection phases at the level just above the cylinders. (d) The distribution of the electric field for the plane wave incident on a disordered magnetic mirror.

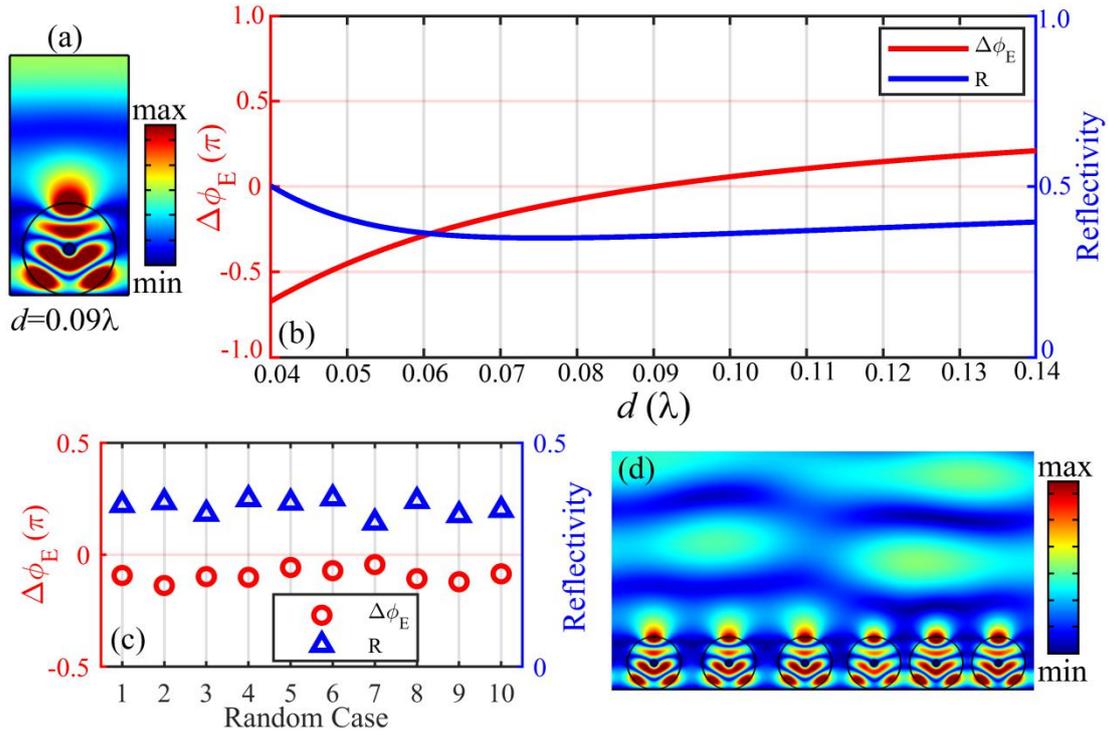

Fig. 4 (a) The distribution of normalized total electric field in a period. The magnetic mirror is in a periodic structure with the spacing between the neighboring cylinders $d = 0.09\lambda$. The ratio of inner radius to outer radius is $\alpha=0.12$. (b) The reflection phase $\Delta\phi_E$ (red solid line) and reflectivity (blue solid line) of the magnetic mirrors in periodic structures as functions of the spacing $d$. (c) The reflectivities (blue triangles) and averaged reflection phases for different disordered samples. The averaged reflection phase is obtained by spatially averaging the reflection phases at the level just above the cylinders. (d) The distribution of the electric field for the plane wave incident on a disordered magnetic mirror.

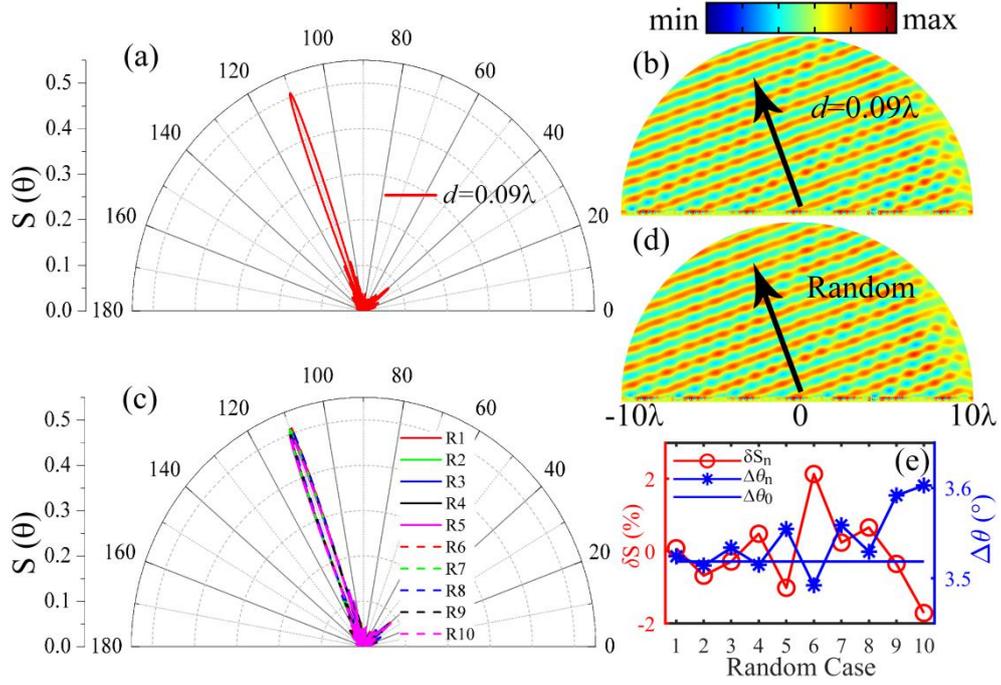

Fig. 5 (a) The polar plot of the reflection intensity S($\theta$) for the metasurface in a periodic structure with spacing $d = 0.09\lambda$. $\alpha$ is ranging from 0.20 to 0.62. (b) The corresponding normalized reflection wave field distribution. The arrow denotes the main propagating direction of the reflection wave. (c) The polar plots of the reflection intensity S($\theta$) for the 10 disordered metasurfaces with nonuniform and random spacing. (d) The normalized reflection wave field distribution of an example disordered metasurface. (e) The relative differences on the peak values of the reflection intensity S($\theta$) between the disordered and ordered metasurfaces (red circle line) and the angle expansions $\Delta\theta_n$ of the main lobes (from $100°$ to $120°$) of S($\theta$) for the ten disordered metasurfaces (blue star line). The angle expansion $\Delta\theta_0$ of the main lobe of S($\theta$) for the ordered metasurface is shown by the blue line.

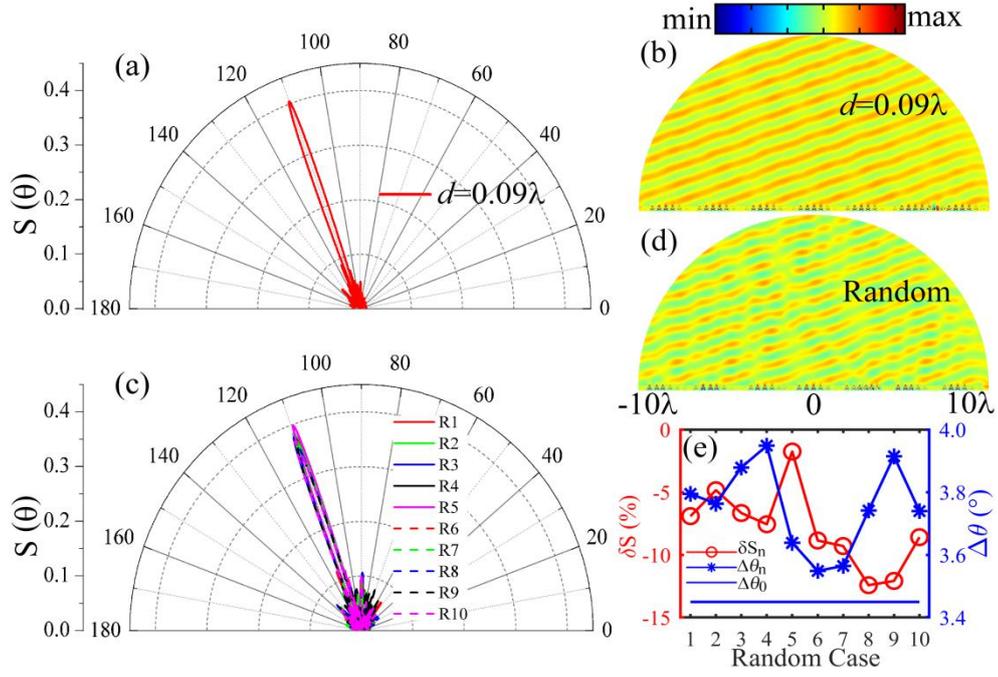

Fig. 6 (a) The polar plot of the reflection intensity S($\theta$) for the metasurface in a periodic structure with spacing $d = 0.09\lambda$. $\alpha$ is ranging from 0.01 to 0.20. (b) The corresponding normalized reflection wave field distribution. The arrow denotes the main propagating direction of the reflection wave. (c) The polar plots of the reflection intensity S($\theta$) for the 10 disordered metasurfaces with nonuniform and random spacing. (d) The normalized reflection wave field distribution of an example disordered metasurface. (e) The relative differences on the peak values of the reflection intensity S($\theta$) between the disordered and ordered metasurfaces (red circle line) and the angle expansions $\Delta\theta_n$ of the main lobes (from $100°$ to $120°$) of S($\theta$) for the ten disordered metasurfaces (blue star line). The angle expansion $\Delta\theta_0$ of the main lobe of S($\theta$) for the ordered metasurface is shown by the blue line.

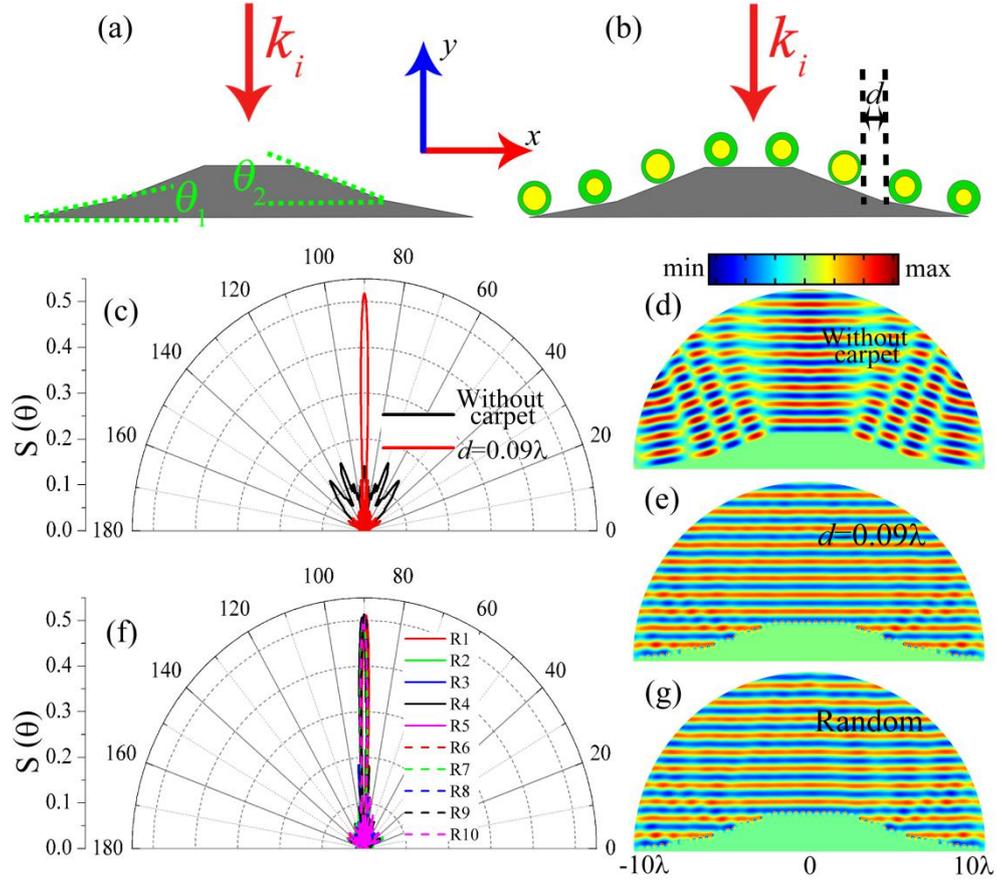

Fig. 7 (a) The geometry of the bump. The bump is symmetric and has two tilt angles. (b) Schematical show of the metasurface on the bump. The horizontal spacing $d$ is defined as the distance between the vertical tangential lines of the neighboring cylinders. (c) Reflection intensities S($\theta$) for the bump without the metasurface covered (black solid line) and the bump with the metasurface covered. The metasurface is composed of cylinders with $a$ ranging from 0.20 to 0.62 the same horizontal spacing $d = 0.09\lambda$ (red solid line). (d) and (e) the total electric field distributions for the uncovered and covered bump. (f) Reflection intensities S($\theta$) for the bump with 10 different disordered metasurfaces covered. (g) The total electric field distribution for the bump with an example disordered metasurface covered.

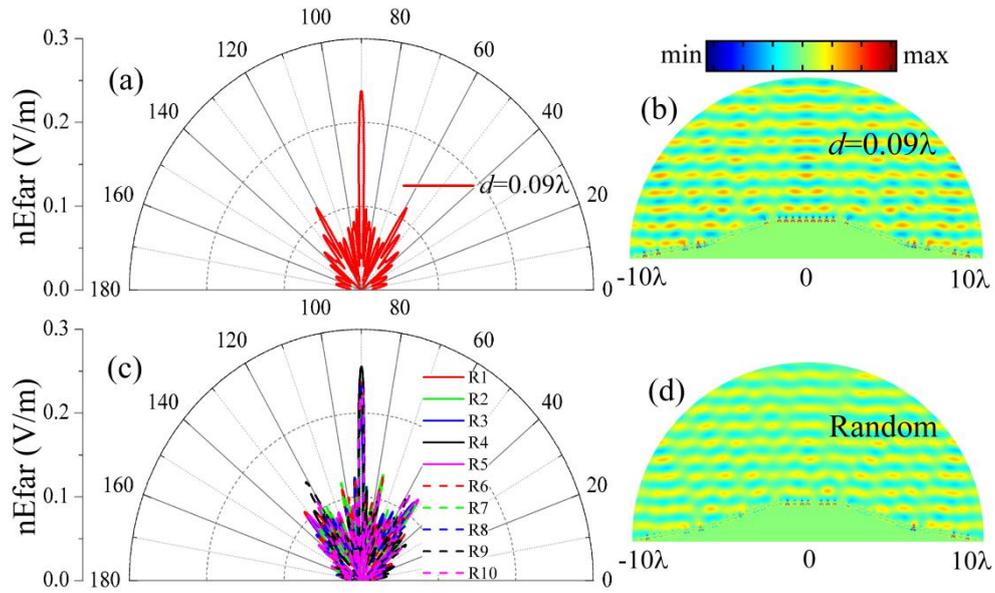

Fig. 8 (a) The reflection intensity S($\theta$) for the bump with the metasurface covered. The metasurface is composed of cylinders with $\alpha$ ranging from 0.01 to 0.20 and the same horizontal spacing $d = 0.09\lambda$. (b) The total electric field distribution for the bump with the ordered metasurface covered. (c) Reflection intensities S($\theta$) for the bump with 10 different disordered metasurfaces covered. (d) The total electric field distribution for the bump with an example disordered metasurface covered.